\documentclass[aps,a4paper,twocolumn,10pt,prl,showpacs]{revtex4-1}
\usepackage{graphicx, amsmath}
\usepackage{hyperref,color}
\def \degree {^\mathrm{o}}

\begin{document}

\title{Ship wakes: Kelvin or Mach angle?}

\author{Marc Rabaud$^1$, Fr\'{e}d\'{e}ric Moisy$^{1,2}$}

\affiliation{$^1$Laboratoire FAST, Universit\'e Paris-Sud, UPMC Universit\'e Paris 6, CNRS. B\^{a}t. 502, Campus universitaire, 91405 Orsay, France \\
$^2$Institut Universitaire de France}

\date{\today}


\begin{abstract}

From the analysis of a set of airborne images of ship wakes, we show that the wake angles decrease as $U^{-1}$ at large velocities, in a way similar to the Mach cone for supersonic airplanes. This previously unnoticed Mach-like regime is in contradiction with the celebrated Kelvin prediction of a constant angle of $19.47\degree$ independent of the ship's speed. We propose here a model, confirmed by
numerical simulations, in which the finite size of the disturbance explains this transition between the Kelvin and Mach regimes at a Froude number $Fr = U/\sqrt{gL} \simeq 0.5$, where $L$ is the hull ship length.

\end{abstract}

\maketitle

The V-shaped wakes behind objects moving on calm water is a fascinating wave phenomenon with important practical implications for the drag force on ships~\cite{Garrett} and for bank erosion along navigable waterways~\cite{Osborne}. The wake pattern was first explained by Lord Kelvin, who by recognizing the dependence of the phase speed $c_\varphi$ of surface gravity waves on their wavelength (dispersion) predicted that the wake half-angle $\alpha_K =\arcsin(1/3)\approx 19.47\degree$ should be independent of the object's velocity $U$~\cite{Lighthill,Darrigol}. However, Kelvin's analysis is called into question by numerous observations of wakes significantly narrower than he predicted~\cite{Reed_2002,Fang_2011,Brown89,Zhu08,Mei91,Munk87}, which have  been rationalized by invoking finite-depth effects~\cite{Fang_2011}, nonlinear resonances or solitons~\cite{Brown89,Zhu08}, unsteady forcing \cite{Mei91}, and visualisation biases~\cite{Munk87}. Analysing a set of airborne images taken from the Google Earth database~\cite{Google}, we show here that ship wakes undergo a transition from the classical Kelvin regime at low speeds to a previously unnoticed high-speed regime $\alpha \sim U^{-1}$ that resembles the Mach cone prediction $\alpha =\arcsin(c_\varphi/U)$ for supersonic airplanes~\cite{Lighthill}.

Since the pioneering work of Froude and Kelvin, waves generated by ships have received considerable interest in naval hydrodynamics, because an important part of the resistance to motion of a ship is due to the energy radiated by these waves~\cite{Darrigol,Garrett,Raphael}. A key parameter governing the wave drag is the hull Froude number, $Fr=U/\sqrt{gL}$, where $L$ is the hull length and $g$ the gravitational acceleration. This non-dimensional number can be conveniently rewritten as $Fr = \sqrt{\lambda_g / 2\pi L}$, where $\lambda_g = 2\pi U^2/g$ is the wavelength of the gravity wave propagating in the ship direction with a phase speed equal to $U$. We propose here a model that takes into account the finite length of the ship, and which successfully predicts the Kelvin-Mach transition at a critical Froude number $Fr = U / \sqrt{gL} \simeq 0.5$.

\begin{figure} [b]
\begin{center}
\includegraphics[width= 6.3 cm]{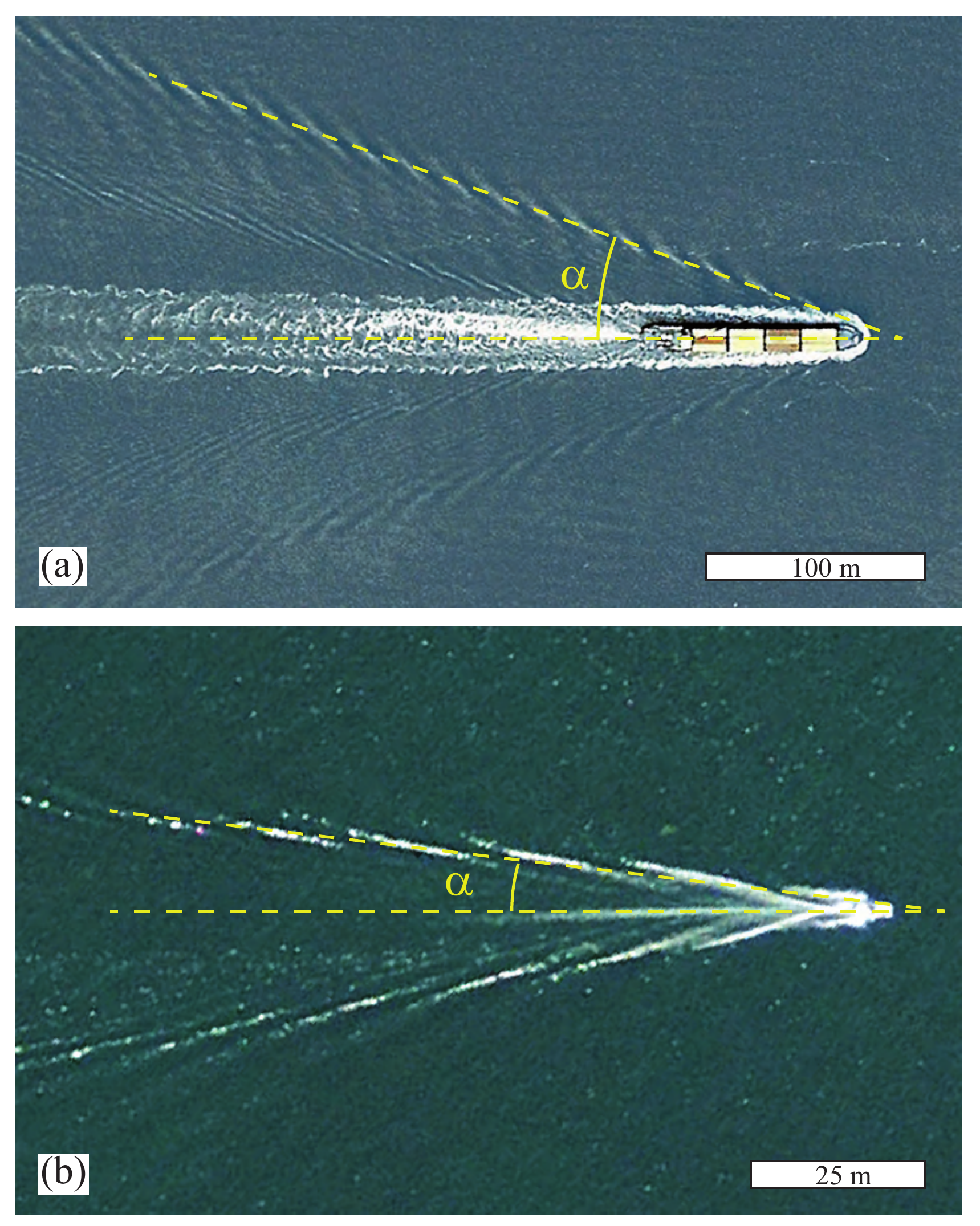}
\caption{(color online) Airborne images of ship wakes taken from the Google Earth database~\cite{sm}. (a), Cargo ship near Antwerpen, with $\alpha \simeq 20\degree$ and Froude number $Fr \simeq 0.15$. (b), Speed boat near Toronto, with $\alpha \simeq 9\degree$ and $Fr \simeq 1.03$. For each image, the wake angle $\alpha$ is defined from the slope of the line going through the brightest points resulting from sun glitter or whitecaps (yellow dotted line), which trace the maximum amplitude of the wake. Using the highest available magnification, the ship length $L$ can be accurately determined with the calibration provided by Google Earth.}
\label{fig:photos}
\end{center}
\end{figure}

We have systematically measured the angle of ship wakes from a series of airborne images taken from the Google Earth database \cite{Google} (data available as Supplemental Material~\cite{sm}). These images, which are corrected for parallax distortion, are chosen close to active harbors, where a high resolution of order of $1$~m is available. Only images where the ship wake forms straight arms, ensuring a constant ship direction, are selected.  For each image, the ship length $L$ is measured and its velocity $U$ is deduced from the wavelength $\lambda$ measured in the wake arms [see Eq.~(\ref{Eq:stationnaire}) below and Ref.~\cite{Aguiar09}], from which the Froude number is determined. 
From these images, wake angles close to the Kelvin prediction $\alpha_K = 19.47\degree$ are systematically found at low $Fr$, i.e. for $L \gg \lambda_g$ (Fig.~\ref{fig:photos}a). In this case, a double wedge pattern, generated by the bow and the stern, can be observed.  At larger $Fr$, $\lambda_g$ reaches the ship length $L$, resulting in interacting bow and stern waves. At this point it is known that the trim of the boat is affected and the wave drag strongly increases: this is the so-called hull limit velocity~\cite{Darrigol,Garrett} --- even if powerful speed boats or sailing boats nowadays overcome this limit. At even larger $Fr$ ($L \ll \lambda_g$), the hull partly rises out of the water, entering in the so-called planing regime~\cite{Garrett}. It is in this large-$Fr$ regime that we find examples of narrow wakes, with angles of $10\degree$ or less, as illustrated in Fig.~\ref{fig:photos}b.

\begin{figure} [b]
\begin{center}
\includegraphics[width= 6.7 cm]{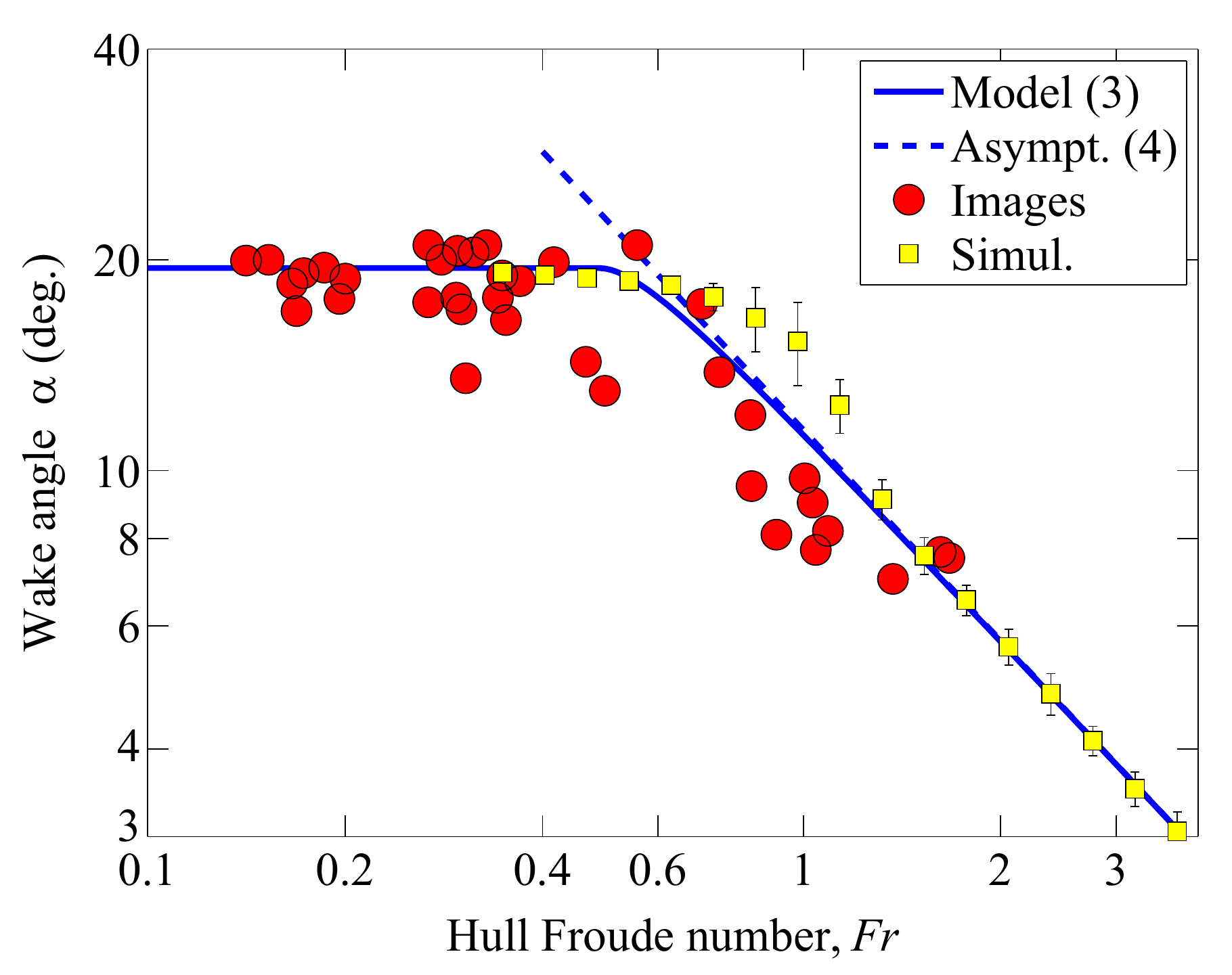}
\caption{(color online) Log-log plot of the wake angle $\alpha$ as a function of the hull Froude number $Fr = U/\sqrt{gL}$. Red circles: angles measured from the 37 airborne images of the dataset~\cite{sm}. Blue line: model (\ref{Eq:alphac}). Blue dotted line: asymptotic law (\ref{Eq:asympt}). Yellow squares: numerical simulations.}
\label{Fig_Google}
\end{center}
\end{figure}

The wake angles measured from the airborne images are plotted as a function of the Froude number in Fig. \ref{Fig_Google}. In spite of a significant scatter, which can be mainly ascribed to the uncertainty in measuring the wavelength $\lambda$, the data clearly shows a plateau at $\alpha \simeq 18.6\degree \pm 1.8\degree$ up to $Fr \simeq 0.5 \pm 0.1$, in good agreement with the Kelvin theory. For larger $Fr$, this Kelvin regime is followed by a decrease of the angle down to values of order of $7\degree$ for the fastest boats of our dataset~\cite{sm}. Interestingly, like in the Mach cone problem, this decrease approximately follows a law $\alpha \sim 1/Fr$.

\begin{figure} [b]
\begin{center}
\includegraphics[width= 6.1 cm]{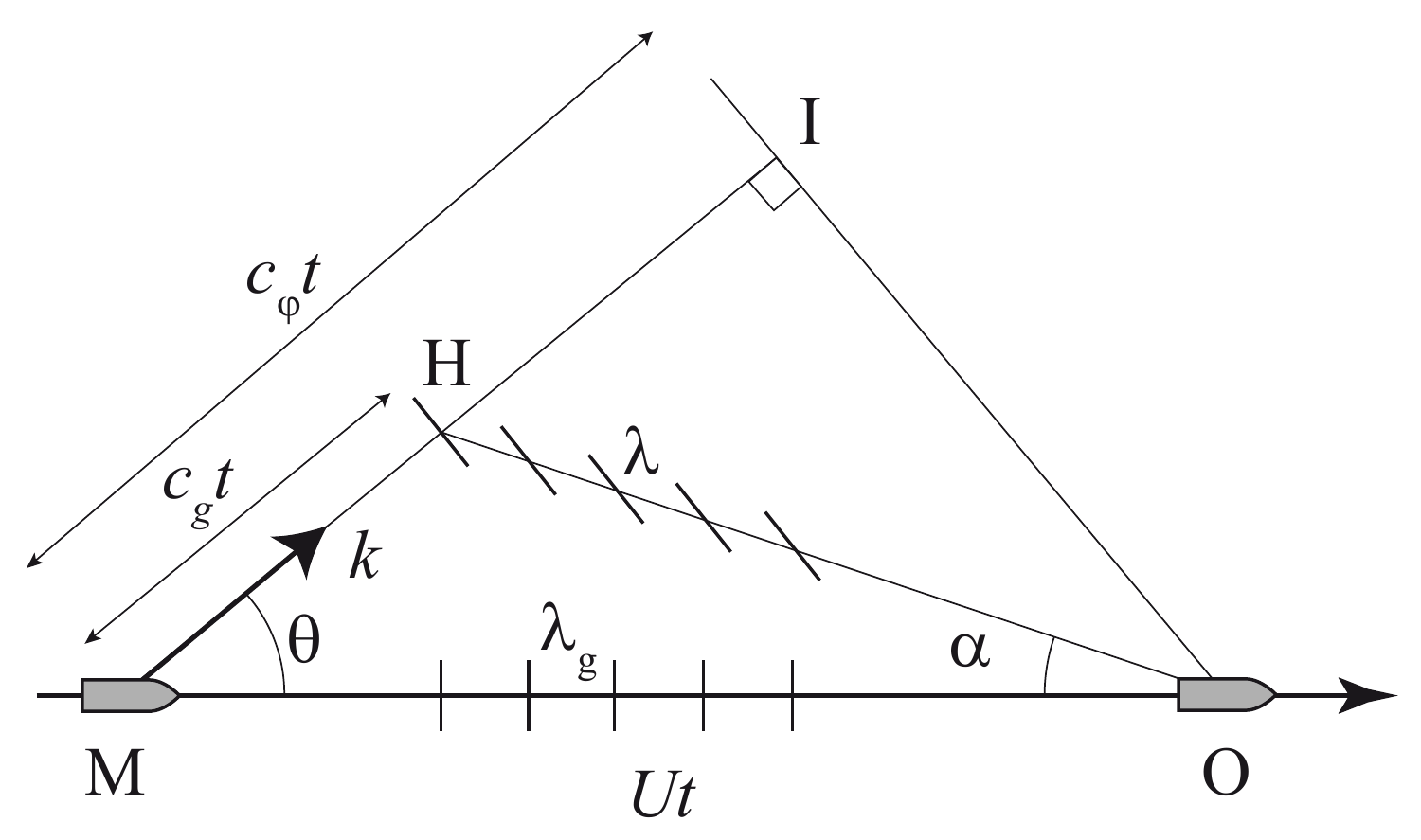}
\caption{Wake angle $\alpha(k)$ of a given Fourier component $k$. The wave of wavenumber $k$ emitted in the direction $\theta(k)$ at time $-t$ when the ship was in M reaches at time 0 the point H at the middle of MI because the group velocity is half the phase speed. The wave crests illustrate the transverse waves $\lambda_g$ at $\alpha=0$ and the divergent waves $\lambda$ at $\alpha \neq 0$ (with $\lambda=2\lambda_g/3$ for the classical Kelvin wake).}
\label{Fig_Craw}
\end{center}
\end{figure}

\begin{figure} [b]
\begin{center}
\includegraphics[width= 6.8 cm]{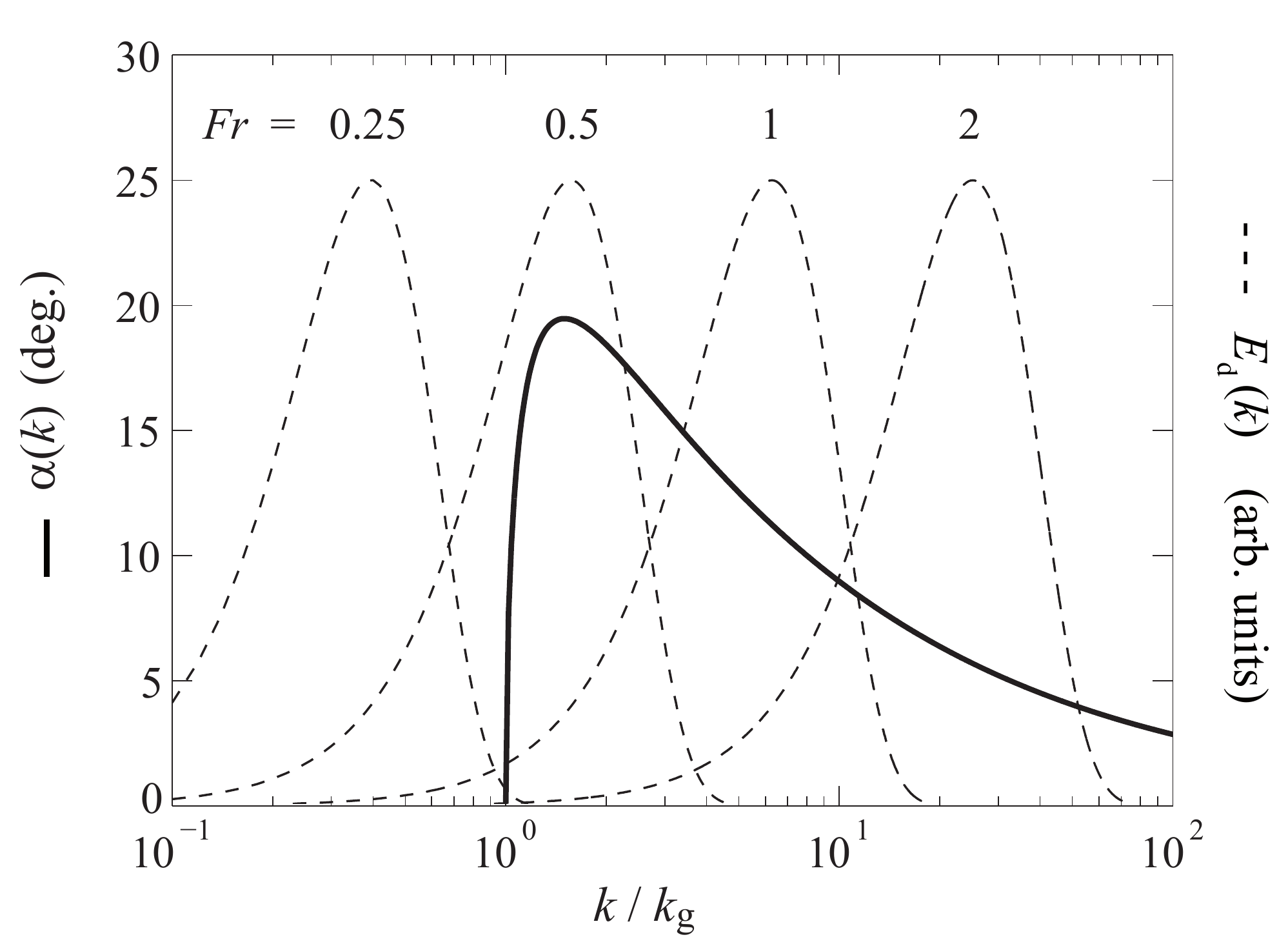}
\caption{Selection of the wake angle by the disturbance spectrum.
Thick line: wake angle $\alpha(k)$ [Eq.~(\ref{Eq:alpha})] as a function of the normalized wavenumber $k/k_g$, with $k_g = g/U^2 = 2\pi / \lambda_g$. Thin dashed lines: disturbance spectrum $E_d(k)$ used in the numerical simulation, plotted for various Froude numbers. The resulting wake angle $\alpha$ is given by the maximum of $\alpha(k)$ taken over the range of significant $E_d(k)$.}
\label{Fig_alphak}
\end{center}
\end{figure}

In order to explain the transition in the wake angle, the starting assumption is to
relate each wavenumber $k$ emitted by the ship hull to a specific angle $\alpha(k)$.
This $k$-dependent angle can be inferred from the linear dispersion relation for gravity waves in deep water~\cite{Lighthill}, $\omega^2 = gk$, from which it follows that the group velocity $c_g = d \omega / d k$ of each wavenumber $k$ is half its phase speed $c_\varphi = \omega/k = \sqrt{g/k}$.  Since the wake is stationary in the reference frame of the ship, the phase speed of each $k$ must be given by the ship velocity projected in the direction of the wave propagation,
\begin{equation}
U\cos \theta(k)=c_\varphi(k)=\sqrt{g/k}.
\label{Eq:stationnaire}
\end{equation}
Accordingly, only wavenumbers $k \geq k_g = 2\pi/\lambda_g = g/U^2$ can form a stationary pattern.
Following the geometrical construction of Ref.~\cite{Crawford_1984} (see Fig.~\ref{Fig_Craw}), we consider a wave of given wavenumber $k$ emitted at time $-t$ in the direction $\theta(k)$ given by Eq.~(\ref{Eq:stationnaire}) when the boat was in M, with MO$=Ut$. Since its group velocity is half its phase speed, the distance MH$=c_g t$ traveled by this wave is half the distance MI$=c_\varphi t$. It follows that the wedge angle formed by this particular wavenumber $k$ is
\begin{equation}
\alpha(k)= \tan^{-1} \frac{\sqrt{k / k_g-1}}{2 k / k_g-1},
\label{Eq:alpha}
\end{equation}
which is plotted in Fig.~\ref{Fig_alphak}. This angle vanishes at the lower bound $k=k_g$ allowed by the ship velocity (corresponding to the transverse wave $\lambda_g$ shown in Fig.~\ref{Fig_Craw}) and at $k \rightarrow \infty$, and reaches the maximum $\alpha_K=\tan^{-1} (1/\sqrt{8}) \simeq 19.47\degree$ at $k = 3 k_g/2$.
No energy can be found outside this wedge of angle $\alpha_K$, so if all wavenumbers are excited (flat disturbance spectrum), the classical Kelvin angle $\alpha_K$ is found.

An important point is that the departure from the Kelvin angle $\alpha_K$ in Fig.~\ref{Fig_Google} is observed at large velocities, for which both viscosity and surface tension effects can be neglected, so a model for the Kelvin-Mach transition must rely only on gravity waves. The key assumption here is that a ship of size $L$ cannot excite waves of size much larger than $L$, suggesting to model the finite size of the ship by a disturbance spectrum $E_d(k)$ truncated at low wavenumber. The simplest choice is a Heaviside step spectrum, $E_d(k) \propto H(k-k_L)$, with a cutoff wavenumber $k_L= 2\pi/L$. The resulting wake angle is therefore simply given by the maximum of (\ref{Eq:alpha}) taken over the range of excited wavenumbers, $[k_L, \infty[$.  When $k_L \leq 3 k_g/2$, i.e. for $Fr \leq Fr_c = \sqrt{3/4\pi} \simeq 0.49$, the maximum of $\alpha(k)$ belongs to the excited range and the classical Kelvin angle is found; on the other hand, when $k_L > 3 k_g/2$, a lower angle given by Eq.~(\ref{Eq:alpha}) at $k=k_L$ is selected, resulting in a piecewise wake angle
\begin{subequations} \label{Eq:alphac}
      \begin{align}
        \alpha &= \tan^{-1} (1/\sqrt{8}) \simeq 19.47\degree, \quad &Fr\leq Fr_c \label{Eq:alphaca}\\
        \alpha &= \tan^{-1} \frac{\sqrt{2\pi Fr^2-1}}{4 \pi Fr^2-1}, \quad &Fr \geq Fr_c, \label{Eq:alphacb}
      \end{align}
\end{subequations}
This model provides a good comparison with the angles measured from the wake images, as shown in Fig.~\ref{Fig_Google}. Values slightly below Eq.~(\ref{Eq:alphacb}) probably originate from a systematic underestimation of the Froude number for the fastest boats: at such large $Fr$ boats are in the planing regime, resulting in a waterline length smaller than their actual length $L$ seen from above.  It is worth noting that the details of the high wave-number part of the disturbance spectrum, which must be affected by complex flow phenomena around the ship hull (flow separation, capillary effects, splashing), is not critical in this model. All disturbance spectrum with no energy below $k_L = 2\pi / L$ would produce essentially the same transition at $Fr=Fr_c$, and would differ only in the limit of low Froude number.

Interestingly, for large Froude numbers, Eq.~(\ref{Eq:alphacb}) simply reduces to
\begin{equation}
\alpha \approx \frac{1}{2 \sqrt{2\pi} Fr},
\label{Eq:asympt}
\end{equation}
which is analogous to the Mach cone angle for (non-dispersive) acoustic waves, $\alpha \simeq c_g / U$, where the constant group velocity $c_g = \frac{1}{2} \sqrt{g/k_L}$ selected by the ship length $L$ plays here the role of the sound velocity. We can therefore call the regimes described by Eq.~(\ref{Eq:alphaca})
and (\ref{Eq:alphacb}) the 'Kelvin' and 'Mach' regimes, respectively. The asymptotic law (\ref{Eq:asympt}) matches the model (\ref{Eq:alphacb}) to within 1\% for $Fr > 1$, and describes equally well the data in Fig.~\ref{Fig_Google}. Note that Eq.~(\ref{Eq:asympt}) is similar to what would be obtained in the case of a (non-dispersive) shallow water wake~\cite{Fang_2011}, namely $\alpha \approx 1/Fr_H$, where $Fr_H = U / \sqrt{gH}$ is now the {\it depth} Froude number based on the sea depth $H$. But despite this resemblance, Eq.~(\ref{Eq:asympt}) remains essentially a {\it dispersive} result, as confirmed by the the characteristic feathered wave pattern seen in Fig.~\ref{fig:photos}(b). Indeed, a {\it non-dispersive} shallow-water wake would be made of two straight crests comparable to a supersonic shock wave, which we have never observed in our set of images. The finite size rather than finite depth origin of the narrow wakes analysed here 
is further confirmed by the depth Froude number determined for each location~\cite{navionics}, which does not correlate to the measured angles~\cite{sm}.

In order to confirm the influence of the finite size of the disturbance on the wake angle, we have performed a pseudo-spectral simulation of the wave pattern generated by a disturbance moving at constant velocity with infinite water depth.
The simulation is carried out
in a square domain ${\bf r} = (x,y) \in [-D/2, D/2]^2$, discretized on a grid of size $N=2048$. At each time step $\delta t = \delta x / U$, where $\delta x = D/N$ is the mesh unit, a disturbance $\delta \zeta$ located at ${\bf r_s} = (D/4, 0)$ is added to the surface deformation $\zeta({\bf r},t)$ initially set to 0. The disturbance is a localized deformation of the interface $\delta \zeta = w_d({\bf r - r_s})\delta t$, mimicking the effect of a pressure disturbance applied during $\delta t$. Since the simulation is performed in the reference frame of the disturbance, the actual deformation field $\zeta({\bf r},t)$ is translated by one mesh unit $\delta x$, Fourier-transformed, and each wave component is phase-shifted according to the dispersion relation, $\hat \zeta({\bf k}, t+\delta t) = \hat \zeta({\bf k}, t) \exp [i \omega({\bf k}) \, \delta t]$.  The resulting spectrum is then Fourier-transformed back in the physical domain, yielding $\zeta({\bf r}, t+\delta t)$, and an absorbing boundary condition is applied in order to avoid the wake pattern re-entering the periodic domain. The procedure is repeated until a stationary pattern is achieved, i.e. when the transient waves generated at $t=0$ leave the domain. Note that the resulting deformation $\zeta({\bf r})$ is complex, with the real part being the actual surface deformation (related to the potential energy), and the imaginary part coding for the phase of the wave (related to the kinetic energy, which is in phase quadrature with the potential energy for each Fourier component).

The deformation disturbance $w_d({\bf r})$ is the response of the free surface to an applied moving pressure distribution $P({\bf r} - U t {\bf e}_x)$, given by $- U \partial P / \partial x$ in the frame of the disturbance. Introducing a highly simplified hull disturbance in the form of an axisymmetric Gaussian pressure distribution $P({\bf r})$, the resulting disturbance deformation has a dipolar shape, which we write as $w_d({\bf r}) \propto - \partial [\exp(-2 \pi^2 ({\bf r}/L)^2)] /\partial x$, with a bump before and a hole behind it. The corresponding disturbance spectrum is $E_d(k) = |\hat w_d({\bf k})|^2 \propto k_x^2 | \hat P({\bf k}) |^2 \propto k_x^2 \exp[- (kL)^2 /4\pi^2]$, which is maximum at $k_x = 2\pi/L$. This model spectrum (plotted in Fig.~\ref{Fig_alphak} for four values of the Froude number) has therefore an effective low-wavenumber cutoff, which is the fundamental ingredient of the Heaviside step spectrum leading to the wake angle (\ref{Eq:alphac}). Its decrease at large $k$ (which is required for numerical convergence) should not affect the wake angle selection, provided that the Froude number is not too small.

\begin{figure} [b]
\begin{center}
\includegraphics[width= 6.4 cm]{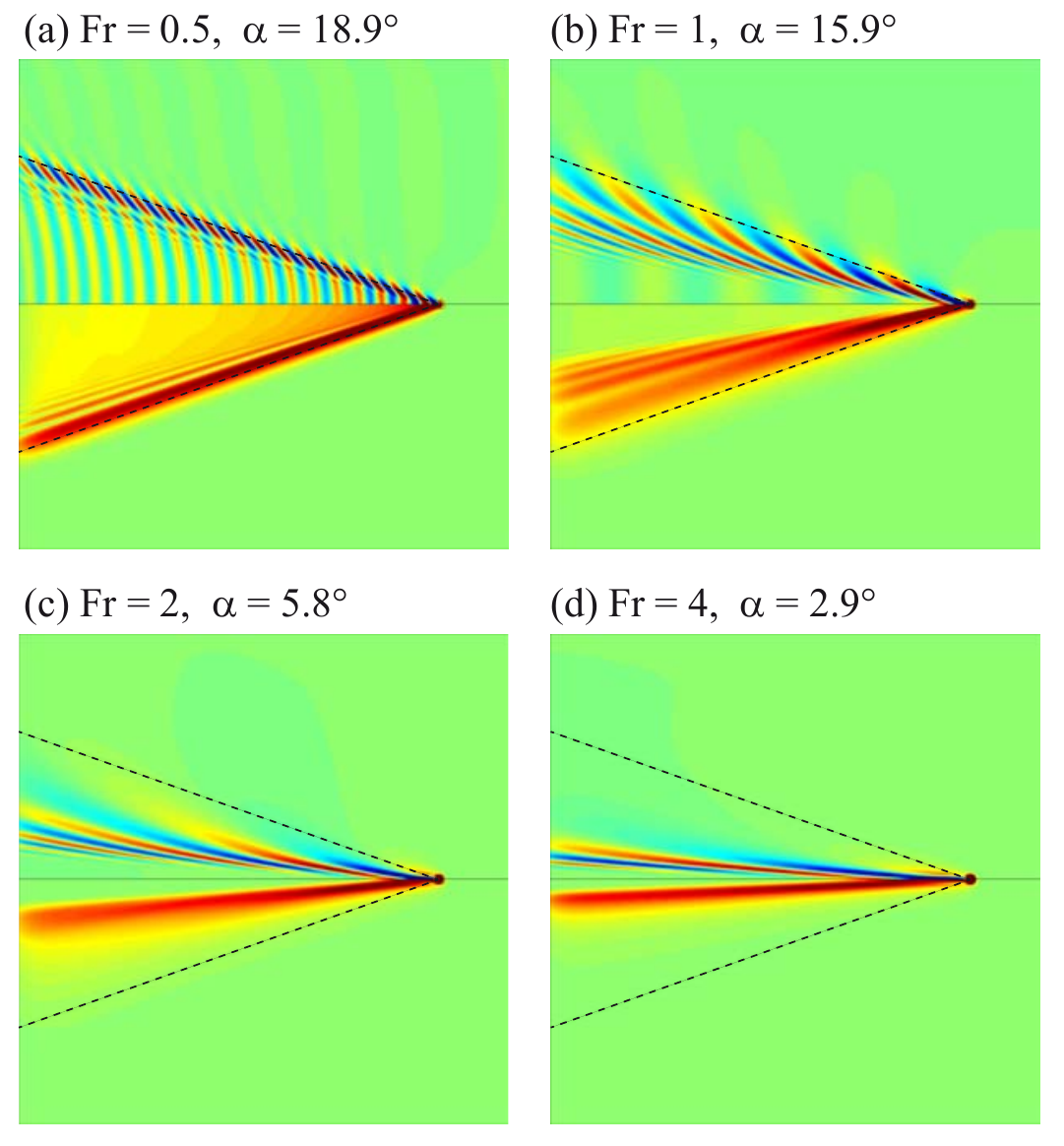}
\caption{(color online) Wake pattern obtained from numerical simulation, for Froude numbers $Fr = 0.5, 1, 2$ and 4. The disturbance size is $L = 4$~m, and the imaged domain is 140~m.  The upper panel of each image shows the physical amplitude ${\cal R} \{ \zeta(x,y) \}$, and the lower panel shows the modulus $|\zeta(x,y)|$. Wake angles (shown as yellow squares in Fig.~\ref{Fig_Google}) are measured from best linear fit of 
the maximum of $|\zeta(x,y)|$. Black dotted lines indicate the Kelvin angle $\alpha_K = 19.47\degree$.}
\label{Fig_Pattern_simule}
\end{center}
\end{figure}

The simulated wake patterns shown in Fig.~\ref{Fig_Pattern_simule} for four Froude numbers reproduce successfully the key features of the ship wake observations. Interestingly, at $Fr=0.5$, the transverse wave of wavelength $\lambda_g$ is clearly present in the field, in addition to the divergent wave of wavelength $2 \lambda_g/3$ along the cusp line at $\alpha \simeq 19\degree$, as commonly observed behind boats at moderate Froude numbers. Froude numbers below 0.3 produce no wake, because the range of wavenumbers $k \in [k_g, \infty[$ is not significantly fed by the disturbance spectrum (see the curve at $Fr=0.25$ in Fig.~\ref{Fig_alphak}); this is a limitation of the smooth pressure distribution chosen here, which does not possess the high-wavenumber energy content of the Heaviside spectrum used in the model. At larger Fr, only the divergent wave is present, since the transverse component at $k_g$ falls outside the disturbance spectrum $E_d(k)$. We can note the broad wake arms at $Fr=1$, for which $E_d(k)$ covers a range where $\alpha(k)$ varies significantly. At larger $Fr$, $E_d(k)$ picks only a narrow range of $\alpha(k)$, and the wake angle becomes more precisely selected. The wake angles have been measured from a linear fit through the maxima of the modulus of the complex deformation $|\zeta(x,y)|$ (shown in Fig.~\ref{Fig_Pattern_simule}), which conveniently displays the square root of the total energy in the physical space.
The resulting wake angles, also plotted in Fig.~\ref{Fig_Google} (yellow squares), are in good agreement with the two branches (\ref{Eq:alphaca}) and (\ref{Eq:alphacb}), confirming that the Kelvin-Mach transition at $Fr \simeq 0.5$ is correctly captured by the finite size effect of the disturbance.

Our results suggest that the departure from conventional Kelvin wakes reported in the literature can be attributed at least in part to the effect of the finite size of the disturbance. The Mach-like ship wakes described here provide an intriguing example of a seemingly non-dispersive wake pattern, similar to a supersonic shock wave, although keeping its specific feathered shape characteristic of a dispersive medium.
We are now extending this approach to smaller scales, e.g. for ducks or insects~\cite{Benzaquen,Rousseaux}, for which richer wake patterns are expected from the interplay between the capillary cutoff and the finite size of the moving body.

\begin{acknowledgments}

We acknowledge N. Pavloff, E. Rapha\"el and G. Rousseaux for fruitful discussions, and J.P. Hulin and N. Ribe for their comments to the manuscript.

\end{acknowledgments}

\end{document}